# Circumneutral concentrated ammonium acetate solution as water-in-salt electrolyte

Mohammad Said El Halimi[a,c], Federico Poli[a], Nicola Mancuso [a], Alessandro Olivieri [a], Edoardo Jun Mattioli[b], Matteo Calvaresi [b], Tarik Chafik [c], Alberto Zanelli [d],

Francesca Soavi*[a,d]

[a] Laboratory of Electrochemistry of Material for Energetics, Department of Chemistry "Giacomo Ciamician", Alma Mater Studiorum Università di Bologna, Bologna 40126, Italy

[b] NanoBio Interface Laboratory, Department of Chemistry "Giacomo Ciamician", Alma Mater Studiorum Università di Bologna, Bologna 40126, Italy

[c] Laboratory of Chemical Engineering and Ressources Valorization, Faculty of Sciences and Techniques of Tangier, University Abdelmalek Essaadi, Tangier 90000, Morocco

[d] Istituto per la Sintesi Organica e la Fotoreattività (ISOF), Consiglio Nazionale delle Ricerche (CNR), Bologna 40129, Italy

*Corresponding author: francesca.soavi@unibo.it

**Abstract**

The exponentially growing market of electrochemical energy storage devices requires substitution of flammable, volatile, and toxic electrolytes. The use of Water in salt solutions (WiSE) regarded as green electrolyte might be of interest thanks to an association of key features such as high safety, low cost, wide electrochemical stability, and high ionic conductivity. Here, we report comprehensive chemical-physical study of circumneutral WiSE based on ammonium acetate so as to investigate application in electrochemical energy storage systems, with focus on the effect of pH, density, viscosity, conductivity, and the ESW with salt concentration ranging from 1 to 30 mol kg$^{-1}$. Data are reported and discussed with respect to the structure of the solutions investigated by complemental IR and molecular dynamic study. The study is addressed through the showcase of an asymmetric supercapacitor based on Argan shell-derived carbon electrodes tested at temperatures ranging from -10 to 80 °C.

1. Introduction

Today, the increasing demand for electrochemical energy storage devices (ESDs) pushes towards improving their performance and safety at lower cost and environmental impact. The electrolyte is a key component of ESDs and should address the following requirements: i) high ionic conductivity, ii) wide electrochemical stability window (ESW), iii) high thermal stability, iv) low cost, and v) environmental compatibility.

Commercial lithium-ion batteries and electrical double layer capacitors (EDLCs) typically feature electrolytes based on lithium hexafluorophosphate in ethylene carbonate and dimethyl carbonate (LP30) and tetraethylammonium tetrafluoroborate in acetonitrile (TEABF$_4$/ACN), respectively. Table 1 summarizes these electrolyte characteristics. The ionic conductivity ($\kappa$) and ESW are 10.8 mS cm$^{-1}$ and 5.7 V for LP30, 56 mS cm$^{-1}$, and 6.1 V for TEABF$_4$/ACN. Their cost is mainly affected by the salt.

Novel alternative electrolytes have been proposed and the main achievements have been excellently reviewed in the literature [1-3]. Ionic liquids are an interesting class of organic electrolytes, indeed besides their good ionic conductivity and electrochemical stability, they present important advantages associated with low vapor pressure and flammability, that are key requisites to design safe ESDs. However, they cannot be considered as totally green and their toxicity has been reported as an issue for the disposal of end-of-life devices [4]. In addition, their high cost still represents a limit to larger exploitation in batteries or supercapacitors. As an example, EMITFSI shows a conductivity of 9 mS cm$^{-1}$ and an ESW of 4.5 V, while for PYR$_{14}$TFSI the conductivity is 2.8 mS cm$^{-1}$ and the ESW is 6.6 V. Their cost is more than 5 times higher than that of TEABF$_4$/ACN and LP30 (Table 1). Hence, the use of aqueous electrolytes offers a promising opportunity to design cheap and safer devices as compared to organic electrolytes because of their non-flammability, low cost, and environmental friendliness. However, due to water splitting conventional aqueous electrolytes place an intrinsic limitation on the ESD, and mainly EDLC, practical cell voltage.

A major breakthrough in electrolytic materials was achieved only a few years ago by increasing the salt concentration in appropriate salt-solvent combinations [14]. The so-called "water in salt electrolytes" (WiSE) are obtained with aqueous solutions containing salt to water volume or mass ratio higher than 1 [10]. Thanks to their molecular structure and water-to-ion interactions, WiSE have been demonstrated to reach unexpectedly wide ESW, beyond the thermodynamic stability limit of water. Therefore, WiSE are receiving considerable attention as safe electrolytes for batteries and supercapacitors [10, 13, 15]. The main WiSE investigated for batteries are based on fluorinated imide-based salts, usually lithium bis (trifluoromethane)-sulfonimide (LiTFSI). Suo et al., were the first to report about a WiSE based on a 21 mol kg$^{-1}$ LiTFSI water system capable of reaching an ESW of 3V and a conductivity of 10 mS cm$^{-1}$ (Table 1) [10]. Since then, the interest in WiSE for lithium-ion batteries and supercapacitors has been growing [13, 15-18]. Although significant improvements were achieved with imide-based WiSE systems, several economic and environmental challenges are still ahead as pointed out by Lukatskaya et al. for LiTFSI [12]. Furthermore, the limited geographical distribution of lithium deposits in the earth's crust, relative to sodium (Na) and potassium (K) deposits, raises another concern associated with the amount of lithium salt needed for WiSE electrolytes [12]. In attempt to lower the amount of lithium salts, binary salts, like eutectic mixtures of lithium and potassium acetates, have been suggested. Mixed WiSE solutions containing 32 mol kg$^{-1}$ potassium acetate – 8 mol kg$^{-1}$ lithium acetate for aqueous batteries featured ESW of 2.7 V and κ of 5.3 mS cm$^{-1}$ [12]. Moreover, EDLCs offer the possibility to us lithium-free WiSE such as potassium acetate–based WiSE, as already reported for an activated carbon-based symmetric supercapacitor that featured excellent cyclic performance under an operating voltage of 2 V [19].

Even though, superconcentrated solutions of acetates are inherently alkaline, due to the hydrolysis reaction of the acetate anion. So far, cheaper sodium perchlorate based WiSE featuring 64.2 mS cm$^{-1}$ and ESW of 2.8 V, has been proposed as a mild neutral electrolyte for 2.3 V EDLCs [13]. Unfortunately, this WiSE cannot be considered as totally green mainly because the perchlorate anion is known as a strong oxidizer [20].

The present work is devoted to the study of safer and less corrosive circumneutral WiSE obtained with a highly concentrated aqueous solution of ammonium acetate (AmAc). This salt features high solubility in water of 1.48 kg L$^{-1}$ and is composed of ions that derive from a weak base and a weak acid with similar pKa and pKb values. Such a particular characteristic makes AmAc solutions circumneutral. Here, a comprehensive chemical-physical study of AmAc solutions with concentrations ranging from 1 to 30 mol kg$^{-1}$ is reported and discussed. Specifically, the trends of pH, density, viscosity, conductivity, and ESW with AmAc concentration will be discussed in terms of solution structure by complementary IR spectroscopy and molecular dynamics (MD) studies. The feasibility of the use of AmAc WiSE in electrochemical energy storage devices will be demonstrated through a showcase of supercapacitors using Argane-shell derived carbon electrodes and AmAc 26.4 mol kg$^{-1}$ electrolyte.

## 2-Materials and Methods

### 2.1 Ammonium acetate WiSE characteristics

Ammonium acetate (98% purity) was purchased from EMSURE. The pH of the aqueous solutions was measured by a pHM210 Standard pHmeter MeterLab. The density of the solutions was obtained by weighting exact volumes of solution (100μL) measured with a P200 micropipette. The viscosity of the solutions was obtained by a Viscoclock SI Analytics bubble viscometer. ATR (Attenuated Total Reflection) spectra of liquid aqueous solutions were carried out with an FTIR Brucker Alpha spectrometer equipped with an ATR head. The limited length of the optical path in the sample eliminated the problem of the strong attenuation of the infrared signal by highly absorbent media such as the aqueous solutions. The ionic conductivity was measured with a CDM210 conductivity meter electrode MeterLab. The solutions were thermostated in a cryostat bath at different temperatures.

*2.2 Molecular dynamics (MD) simulations*

Setting molecular dynamics (MD) simulations. To investigate the behaviour of AmAc solutions at the atomistic level, MD simulations were carried out. Boxes with different AmAc/$H_2O$ ratios (corresponding with the experimental WiSE concentrations investigated in this work) were built. The FF14SB force field was used to model acetate anions and ammonium cations [21] while water molecules were simulated by using the TIP5P water model [22]. Minimization and equilibration. About 5000 steps of steepest descent minimization, followed by additional 5000 steps of conjugate gradient minimization were performed with PMEMD [23]. The minimized structure was considered for a 3step equilibration protocol. Particle Mesh Ewald summation was used throughout and H-atoms were considered by the SHAKE algorithm [23]. A time step of 2 fs was applied in all MD runs. Individual equilibration steps included (i) 50 ps of heating to 298 K within an NVT ensemble and temperature coupling according to Berendsen. (ii) 50 ps of equilibration MD at 298 K to switch from NVT to NPT and adjust the simulation box. Isotropic position scaling was used at default conditions. (iii) 400 ps of continued equilibration MD at 298 K for an NPT ensemble switching to temperature coupling according to Andersen.

MD simulation was carried out for the equilibrated system using PMEMD [23] Simulation conditions were identical to the final equilibrium step (iii). Overall sampling time was 100 ns.

Snapshot structures were saved into individual trajectory files every 1000 time steps, i.e. every 2 ps of molecular dynamics.

Trajectories obtained from MD simulations were post-processed using CPPTRAJ [23,24]. For each simulated box, the density of the solution, the diffusion constants of water and ions, and the radial distribution function g(r) of acetate and ammonium ions were calculated.

*2.3 Supercapacitor electrode preparation*

Agriculture waste-derived carbon was used as electrode material. The detail of the experimental preparation of this carbon was reported in previous work [25]. Briefly, Argan shells collected from the southern region of Morocco were carbonized under $N_2$ flow at 700 °C for 1 h, in order to decompose the organic materials. The obtained material was activated with potassium hydroxide (KOH) using physical mixing process. Practically KOH beads were mixed with carbon using an activating agent-to-carbonized carbon mass ratio of 4:1. The mixture was heat-treated in a muffle furnace for 1h at 850 °C and under N2 atmosphere; the furnace cooled to room temperature. Finally, the sample was washed with HCl (5 M) solution, rinsed with distilled water until reaching neutral pH then dried at 110 °C overnight. The obtained sample named AC-K-PM yields to BET specific surface area of 1937 $m^2$ $g^{-1}$. The full chemical-physical characterization of this sample is reported in [25]. The supercapacitor electrode was obtained using ARG-K-PM and polytetrafuoroethylene (PTFE) as binder (60% suspension in water), as well as multi-walled carbon nanotubes MW-CNT in an 80/10/10 weight ratio, respectively, and dried at 120°C overnight. Finally, an amount of 6.5 mg of this mixture was pressed onto a titanium grid disk of 8 mm diameter to prepare electrodes that were subsequently dried.

*2.4 Electrochemical measurements*

All the electrochemical measurements were performed by a Bio-Logic VSP300 potentiostat/galvanostat. The ESW was evaluated by voltammetric measurements (LSV, Linear Sweep Voltammetry). The working electrode was a glassy carbon electrode (0.07 cm$^{-2}$), a titanium grid (1 cm$^2$), or an aluminum foil (1 cm$^2$). Metal electrodes were used as received. Metal electrodes were used as received. The reference electrode was a saturated calomel electrode (SCE). The counter electrode was a Pt wire.

The supercapacitor was assembled by coupling two identical electrodes impregnated with a 26.4 mol kg$^{-1}$. AmAc electrolyte for 48 h before the electrochemical measurements using a glass fiber filter (Whatman) as a separator. The electrochemical performance of the supercapacitor was evaluated by a two-electrode setup. Cyclic voltammetry (CV), galvanostatic charge-discharge (GCD), and Electrochemical Impedance Spectroscopy (EIS) of the cell were performed at room temperature, -10 °C and 80 °C.

The supercapacitor specific capacitance ($C_{sp}$) was calculated from GCD curves using the following equation [26, 27].

$$C_{sp} = \frac{I \cdot dt}{m \cdot dV} \tag{1}$$

Where I is the discharge current (A), dV/dt is the slope of the discharge curve, and m is the total mass of the two electrodes (in g).

The maximum specific energy (E$_{max}$) and power (Pmax) were determined by applying equation (2) and (3), respectively [28, 29].

$$E_{max}(Wh\ kg^{-1}) = \frac{C_{sp}\ (F\ g^{-1}) \cdot V_{max}^2(V)}{2 \cdot 3.6} \qquad (2)$$

$$P_{max}(W\ kg^{-1}) = \frac{1}{4} \cdot \frac{V_{max}^2(V)}{ESR\ (\Omega) \cdot m\ (kg)} \qquad (3)$$

Where $V_{max}$ is the is the maximum cell voltage and ESR is the cell equivalent series evaluated by the ohmic drop ($\Delta V$) measured at the beginning of discharge (eq. 4). Given that the same current was used for the charge and discharge,

$$ESR = \frac{1}{2} \cdot \frac{\Delta V}{I} \qquad (4)$$

Practical specific energy (E) and power (P) delivered at different current densities were evaluated by the analyses of the discharge profiles by eqs. (5) and (6)

$$E\ (Wh\ kg^{-1}) = I \cdot \int V \frac{dt}{3.6\ m} \qquad (5)$$

$$P\ (W\ kg^{-1}) = 3.6 \frac{E}{\Delta t} \qquad (6)$$

where $\Delta t$ is the discharge time in seconds.

## 3 Results and discussion

*3.1 Physicochemical studies of Ammonium acetate WiSE*

Table 2 reports the acronym of the ammonium acetate solutions that were investigated with the corresponding values of molality, salt to solvent molar ratio, molarity, and density. We investigated solutions containing ammonium acetate (AmAc), with a molality of molality of 1, 5, 10, 15, 20, 26.4 and 30 mol kg$^{-1}$. It is worth nothing that at the highest concentration only two moles of water are sheared every ion of ammonia $NH_4^+$. Table 2 also reports the values of the density (d) that have been calculated under the hypothesis that the molar volumes of AmAc and water are additives (eqs. 1-2 in supplementary information). The experimental results differ only by less than 0.1 % from the calculated values.

To highlight this small difference, we evaluated the excess molar volume EV (Tables 2) i.e., the difference between the experimental molar volume of the solution and the value obtained by considering that salt and solvent molar volumes are additives (eq. 3 in supplementary information). For concentrations lower than 10 m, EV is slightly negative, therefore indicating that a weak volume contraction takes place during the dissolution of AmAc in water. On the contrary, when the concentration rises above 10 m, EV increases up to 0.23 mL at 26.4 m. This relatively positive volume change can be explained by strong ionic and molecular interactions of AmAc ions and water molecules. Specifically, volume expansion could be related to the directional character of hydrogen bonding [30].

Given that our aim was to propose a neutral WiSE, we checked the pH of the different solutions within the concentration range from 1 m to 30 m (Fig 1a) AmAc is composed of weak acidic and base ions that feature the same base and acid constants. Therefore, as demonstrated in section 3 of the Supplementary Information, the pH of ammonium acetate solutions should not change with the salt concentration and be equal to 7. Different from what is expected, Figure 1a shows that the pH values change from almost neutral (1m solution) to slightly basic along with the increase of the concentration. This is apparently attributed to the decrease of the proton activity.

The viscosity of the solution increases almost exponentially with the solution molality, as shown in Fig. 1b. This trend deviates from the linear curve expected for the diluted solution (Einstein equation) due to involved ions interactions [31]. The absence of minimum in the curve excludes the so-called "water structure breaking" associated with the solution ionic field. At the opposite, it indicates that ammonium and acetate ions are strongly hydrated and contribute to a sort of molecular order in solution [32, 33].

It is known that Stokes-Einstein relation relates conductivity to viscosity [31]. In turn, ionic conductivity and temperature are usually described by an Arrhenius relation. The latter applies for solutions involving no-cooperative mechanism for ion conduction. Under this condition, the logarithm of the specific conductivity ($\kappa$) linearly decreases with the reciprocal of the temperature (see section 4 in the Supplementary Information) [34]. Fig. 2a present the Arrhenius conductivity plots of the different solutions. A higher conductivity is achieved at increased temperature, which is due to a decreased viscosity and increased ion mobility. The highest conductivity is obtained with the 5 m solutions, while the 30 m shows the lowest value in the whole temperature range. This finding agrees with the viscosity trends with ammonium acetate concentration discussed above. Note that the conductivity of the most diluted solution (1 m) is in the same order of magnitude as those featured by the most concentrated ones (from 20 m to 30 m). Only the 1 m solution features a clear Arrhenius-like linear plot. When the concentration increases above 1 m, the plots deviate from linearity. This non-Arrhenius behavior has already been reported for ionic liquid electrolytes, and described by the Vogel-Tamman-Fulcher (VTF) eq. (7) [35]

$$k = A_0 \exp\left(\frac{-B}{T - T_0}\right) \tag{7}$$

where T is the absolute temperature, and $A_0$, B, and $T_0$ are adjustable parameters. According to the VTF model, ion diffusivity and conductivity are affected by several processes like molecular. dissociation and cooperative motions. Particularly, the diffusivity is directly related to fluidity (the reciprocal of viscosity) and decreases with the increase of cooperativity. At the contrary, the molar conductivity is positively affected by cooperative processes [36].

Fig 2b plots the variation of the molar conductivity ($\Lambda m$) and fluidity versus molality of the AmAc solutions. In the concentration range from 1m to 10 m, the molar conductivity follows the decrease of fluidity, with almost the same trend. Instead, for concentrations higher than 10m, the conductivity decrease is less marked than that of fluidity. This suggests that molecular dissociation and cooperative process affects ion conductivity of WiSE.

*3.2 Molecular dynamics (MD) simulations and Fourier transform infrared FTIR*

The structure of the solutions was investigated using MD simulations. MD trajectories were used to calculate the solution densities as a function of the solution molality. (Table 2). The computed and experimental densities present the same trend with a higher increase in density values at lower concentrations.

The diffusion constant D of water, acetate anion, and ammonium cation were calculated using the Einstein relation, on the calculated MD trajectories (as implemented in Almabert [23])

$$2nD = \frac{MSD}{t} \tag{8}$$

where n is the number of dimensions, MSD is the mean square displacement. Salt concentration influences the diffusion constant of water. Increasing the salt concentration, the water molecules move slower due to their interactions with ions (Table 3 and Fig. 3). This is the consequence of increased interaction between water and AmAc, which is also reflected by the contraction of the system volume or higher density. Note also, the decreased ion mobility when molality increases, in agreement with the molar conductivity data reported in Fig. 2b. To have an atomistic insight into the structure of the mixtures, atomic radial distribution functions (g(r)) were calculated. In Fig. 4 g(r) functions for the 1 m and 30 m solutions are reported. The Figure shows the cation-water, cation-cation and cation-anion and the anion water, anion-cation and anion-anion radial distribution functions. Ammonium and acetate ions induce long-range electrostatic interactions in the mixture that goes from a "ion in water" behavior (1 m) to a "ionic liquid-like" behavior (30 m). Furthermore, for the 30 m solution, the ammonium and the acetate are strongly hydrated as supposed by the absence of minimum in the curve of Fig 1b that excludes a "structure breaking" effect of ions in the solution Hence, MD simulations clearly support the "structure effect" of ions suggested by the viscosity experimental data.

The FTIR characterization of the solutions was carried out to evaluate specific interactions between ions and water molecules. Fig 5 shows the superimposed IR spectra of the solutions and shows a clear trend in the evolution of the signals. Typical vibrations associated with water are the stretching of the OH bond in the area between 3000 and 3200 cm$^{-1}$. This region is shared with the stretching of the undissociated acetic acid present in solution. The bending of the HOH angle is around 1640 cm$^{-1}$. These peaks strongly characterize the typical 1m aqueous solution and cover the signals referable to the pure salt [37]. On the contrary, the spectrum of the 30 m solution does not reveal interference attributed to the presence of water, looking similar to what is expected for the pure salt. It is important to empathize that, changing solution from 1 m to 30 m, the intensity of the peak associated with HOH bending decreases and becomes insufficiently resolved, thus remaining indistinguishable from the signal relative to the carboxylate ion (symmetrical stretching around 1540 cm$^{-1}$, asymmetric stretching around at 1400 cm$^{-1}$ for the 30 m solution). The marked change in the shape of the spectrum with the gradual emergence of peaks related to well-defined salt can be interpreted as an important index of the increase in salt activity in solution, combined with a marked decrease in water activity. The peak attributable to O-H stretching alone tends to shift towards lower wavenumbers and change shape from a single very large peak to a system of two peaks located around 3190 cm-1 and 3012 cm-1. This phenomenon is probably associated with high salt concentration and possible peaks overlapping associated with the stretching of the N-H bond, occurring approximately in the same region as of O-H bonding. Simultaneously also the C=O asymmetrical stretching around 1635 cm$^{-1}$, and C-O stretching around 1014 cm$^{-1}$, can be appreciated.

Overall, a careful analysis of the evolution of the spectra indicates that for all the signals a certain degree of red peak shift to lower wavenumbers occurs as the salt concentration increases. Indeed, the O-H stretching shifted from 3670-3805 cm$^{-1}$ to 3615-2455 cm$^{-1}$, N-H bending from 1545 cm$^{-1}$ to 1540 cm$^{-1}$, and C=O symmetrical stretching from 1411 cm$^{-1}$ to 1395 cm$^{-1}$. This is apparently associated with the increased hydrogen bond strength involving ions and water molecules yielding to an evolution in the structure of the solution, which, is consistent with the aforementioned pH, density, and viscosity trends.

*3.3-Electrochemical measurement*

The electrochemical stability of AmAc solutions was evaluated by linear sweep voltammetry (LSV) carried out at 20 mV s-1 in 1 m and 26.4 m solutions at glassy carbon electrode (GC) (Fig 6a). As shows, the cathodic limit for the superconcentrated solution is -1.5 V vs SCE. It is much lower than the limit expected for hydrogen evolution via H$^+$. reduction in acidic solution

(ca, 0.242 V vs SCE). The low cathodic limit of the 26.4 m solution is in line with the low $H^+$ concentration (pH≈8) and with the low availability of $H^+$ ions apparently involved in hydrogen bonds with the other ionic species in solution. Considering the anodic limit, it is at around +1.5 V vs SCE. It slightly decreases with the increase of concentration probably due to higher concentration of acetate ions whose oxidation limits anodic stability.

The most interesting aspect is that using WiSEs based on ammonium acetate it is possible to obtain ESW of 2.9 V at GC, higher than 0.4 V compared to typical aqueous solutions, and close to the performance of electrolytes based on organic solvents. In order to evaluate the feasibility of the use of WiSEs in practical devices, ESW was also evaluated at current collectors that are typically used in supercapacitors and batteries, namely titanium grid and aluminum foil. Fig. 6b. compares the LSVs carried out at 20 mV $s^{-1}$ in 26.4m AmAc. With titanium and aluminum, the cathodic limit becomes more positive because these metals promote fast kinetics hydrogen evolution compared to GC. The cathodic limits are -1.07 V and -0.80V vs. SCE with aluminum and titanium, respectively. Alike GC, aluminum features an anodic limit of +1.40 V vs. SCE. On the contrary, for titanium it increases to +2.60 V vs SCE. Such a wide anodic range of titanium has been already observed in LiTFSI-based WISE and attributed to the formation of a surface Ti-oxide film that partially passivates the grids and hinders electrolyte decomposition [38,39]. Accordingly, using titanium and 26.4 m AmAc an outstanding ESW of 3.4 V should be feasible.

GCE was coated by Argan shells derived carbon (ARG-K-PM) and tested by CV using 1m and 26.4 m AmAc at 20 mV $s^{-1}$ (Fig. 6c). Unexpectedly, when ARG-K-PM electrodes are used, the ESW width does not change with the increase of AmAc concentration. Furthermore, the ESW is significantly narrower than what was observed with the titanium and GC electrodes. Indeed, with ARG-K-PM, the ESW is about 1.3 V, with cathodic and anodic limits that can be set at ca. -0.8 V vs SCE and 0.5 V vs SCE, respectively. Indeed, the high surface area of the ARG-

K-PM carbon (1937 m$^2$ g$^{-1}$) enhances the faradic currents related to electrolyte decomposition and narrows the potential ranges available for the supercapacitor electrode charge. This highlights the importance of the evaluation of electrolyte ESW by adopting the same electrodes that will be exploited in energy storage devices.

On the other hand, Fig. 6c even demonstrates that ARG-K-PM electrodes feature an excellent capacitive response, both in 1m and 26.4 m AmAc, that is of ca. 300 F g$^{-1}$. This value has been extracted from the slope of the plot of the voltammetric specific charge vs. electrode potential. The electrochemical preliminary tests of supercapacitors were carried out by a cell with ARG-K-PM electrodes featuring titanium grids and 26.4 m AmAc WiSE. Figure 6c shows that with ARG-K-PM electrodes the cathodic stability range is 2 fold wider than the anodic one. Therefore, to fully exploit the WiSE electrochemical stability window, we adopted an asymmetric configuration of supercapacitor with positive to negative electrode mass loading ratio equal to ca 2 [41]. Taking into account the good conductivity response at low and high temperatures of 26.4 m AmAc, we evaluated the supercapacitor performance at room temperature (RT), -10°C, and 80°C. Fig.7 reports the CV, GCD, EIS, and Ragone plots of the asymmetric supercapacitor. The highest charge cut-off voltage of the symmetric supercapacitor that enabled high coulombic efficiency (> 99 %) was 1.2 V and higher than that feasible with the symmetric device (0.8 V). For comparison, the full electrochemical characterization of a symmetric supercapacitor with two identical ARG-K-PM electrodes and 26.4 m AmAc WiSE, that featured a maximum cell voltage of 0.8 V, is given as supplementary material (Figure S1 and Section 5 of the Supplementary Information).

Fig. 7a shows the CVs at RT carried out with increasing the scan rate from 5 to 50 mV s$^{-1}$ the curves exhibit a quasi-rectangular shape profile demonstrating good capacitive behaviors of the electrodes even at the highest scan rate. The GCD was performed at current density ranging from 0.1 Ag$^{-1}$ to 1 A g$^{-1}$. The GCD profiles at room temperature are reported in Fig. 7b. They exhibit triangular shape indicating good reversibility and capacitive behavior of the device. Also, all GCD curves show a small ohmic drop, therefore suggesting a low ESR. Figures 7c, 7d and 7e compare the CVs (at 10 mVs$^{-1}$), the GCD profiles (at 0.1 A g$^{-1}$) and the Nyquist plots (100 kHz - 10 mHz frequency range) collected at -10°C, RT and 80 °C. As expected, the CV currents in Fig.7c increase with temperature, due to the higher mobility of AmAc ions. A broad peak appears above 0.9 V at 80°C. The specific supercapacitor capacitances from the CV curves in Fig. 7c were 31 Fg$^{-1}$, 46 F g$^{-1}$ and 71 F g$^{-1}$ at -10°C, RT and 80°C. These values correspond to electrode specific capacitance values of 116 F g$^{-1}$, 173 F g-1, and 266 F g$^{-1}$ of ARG-K-PM. The highest specific supercapacitor capacitances were obtained at 0.2 A g-1 (Fig 7d) and resulted 35 F g$^{-1}$, 50 F g$^{-1}$ and 98 F g$^{-1}$ at -10°C, RT and 80°C. Correspondingly, the maximum energy densities $E_{max}$ were 7 Wh kg$^{-1}$ (-10°C), 10 Wh kg$^{-1}$ (RT) and 20 Wh kg$^{-1}$ (80°C). The ESR values evaluated by the ohmic drop at the beginning of discharge resulted in ca. 8 $\Omega\ cm^2$ at 10°C, 4.5 $\Omega\ cm^2$ at RT, and 2.9 $\Omega\ cm^2$ at 80°C. These values well compares with the medium low frequency resistance of the cells shown by the Nyquist plots reported n Fig. 7. It is worth noting the low ESR exhibited by the cells even at the lowest temperature. The plots indicate that the decrease of temperature mainly impacts ion diffusion in the porous electrode architecture (low frequency tail of the Nyquist plots). On the other hand, MD simulation and experimental data reported in the previous sections already indicated that cooperative mechanisms are responsible for AmAc WiSE ion conductivity. In turn, this affects the kinetics of the electrical double layer formation at the electrode/electrolyte interface, especially at the lowest temperatures. From ESR, maximum power densities $P_{max}$ of 3.7 kW kg$^{-1}$ (-10°C), 6.7 kW kg$^{-1}$ (RT), and 10.4 kW kg$^{-1}$ (80°C) were measured.

The practical specific energy and power delivered by the supercapacitor at different currents and temperatures are compared in the Ragone plot reported in Figure 7f. The maximum specific energy is delivered at the lowest current, while the maximum power is featured at the highest current. At 0.1 A g$^{-1}$, the specific energy is 5.9 Wh kg$^{-1}$ (-10°C), 9.2 Wh kg$^{-1}$(RT), and 15.6 Wh kg$^{-1}$ (80°C). At 1 Ag$^{-1}$, the specific power is 350 W kg$^{-1}$(-10°C), 450 W kg$^{-1}$ (RT) and 507 W kg$^{-1}$ (80°C).

Finally, Fig 8a reports the results of a cycle stability test carried out at different current densities, 0.1 A g$^{-1}$ and 1 A g$^{-1}$ at RT and -10°C. For a comparison, Fig 8b reports the trend of the capacitance vs. cycle number of an analogous device that was assembled with the diluted electrolyte 1m AmAc. The two cells featured very good capacitance retention with coulombic efficiencies approaching 100 %. Only by the use of the superconcentrated electrolyte, it was possible to operate the cell at -10°C over a period of four days. The test was also performed at 80°C with a lower $V_{max}$ of 1 V and the results are reported in figure S2. Even at this high temperature, the coulombic efficiency was higher than 98%. However, in this extreme condition, the capacitance faded by 15% in 1000 cycles, because the cell was not hermetically sealed and vapor leakage occurred.

## 4 Conclusion

The low-cost super-concentrated aqueous solutions based on ammonium acetate feature circumneutral pH (pH = 7-8) and ionic conductivity comparable to or higher than typical organic electrolytes. MD simulations confirmed all the experimental results and provided an atomistic picture of the system. The change in the structure of concentrated solutions is due to strong interactions between ions and/or water molecules through the formation of hydrogen bonding that cause an increase in pH values and a decrease in ions mobility.

Ammonium and acetate are strongly hydrated as suggested by the absence of minimum in the curve excluding a "structure breaking" effect of ions in the solution, meaning that there is no destruction in the structure of water by the ionic field, in agreement with the viscosity results.

In turn, the presence of cooperative motions is suggested by the conductivity temperature dependence, that follows a non-Arrhenius behavior like ionic-liquid electrolytes. Moreover, the MD simulation suggested. that mixture goes from an "ion in water" (conventional solutions) to an "ionic-liquid-like" (concentrated solutions) behavior. One of the most interesting aspects is that the WiSE based on 26.4 m exhibits an ESW of 2.22 V at Al foil, 2.9 V at GC, and an outstanding value of 3.4 V when Ti grid was used. Despite such interesting results, the ESW evaluated using ARG-K-PM electrodes was only 1.3 V wide and affected by the high carbon surface area which promoted electrochemical decomposition of the electrolyte. While this finding strongly suggests that ESW is dependent on the kind of electrodes used for the test, it also prevented the development of symmetric supercapacitors with the high cell voltage expected by the study carried out with GC and metal grids.

On the other hand, the ARG-K-PM electrodes obtained by pyrolysis and activation of argan shells exhibited an exceptional specific capacitance of 300 $Fg^{-1}$ in the super-concentrated electrolyte. The asymmetric supercapacitor assembled with ARG-K-PM electrodes and 26.4 m AmAc WiSE was able to operate at 1.2 V, from -10°C to 80°C with outstanding specific capacitance and low resistance. The symmetric cell delivered noticeable specific energy at extreme temperatures and ranged from 5.9 Wh $kg^{-1}$ at -10°C to 15.6 Wh $kg^{-1}$ at 80°C, values that are competitive with those of commercial supercapacitors featuring organic electrolyte. A promising option to explore is the increase of energy performance through AmAc WiSE in hybrid supercapacitor involving redox or pseudocapacitive electrodes that features faradaic process within the WiSE ESW.

Overall, our study suggests that AmAc WiSE deserves consideration as cheap, circumneutral and environmentally friendly alternative electrolytes for designing green energy storage systems.

**Acknowledgements**

This research was funded by the Italy-South Africa joint Research Progra 2018–2020 (Italian Ministers of Foreign Affairs and of the Environments). The bilateral project CNR Italy CNRST Morocco "Green Supercapacitors" (SAC.AD002.014, n. 7974, C.U.P. B54I20000790005) and PPR2 16/17 project CNRST Morocco are also acknowledged. This project has received funding from the European Union's Horizon 2020 research and innovation programme under grant agreement No 963550.

**Table 1.** Ionic conductivity (κ), electrochemical stability window (ESW) at room temperature and commercial cost of different electrolyte solutions and components used in ESDs.

| Electrolyte solution components and composition | Salt Concentration | κ | ESW | Costs | Ref. |
|---|---|---|---|---|---|
| **Conventional Organic electrolytes** | | | | | |
| Tetraethylammonium tetrafluoroborate in Acetonitrile (ACN/TEABF$_4$) | 1 mol/L | 56 mS/cm | 6.1 V | 3.69\$/g$^a$ 130\$/L$^b$ | [5, 6] |
| Lithium hexafluorophosphate in ethylene carbonate and dimethyl carbonate (LP30) (1.0 M LiPF$_6$ in EC/DMC=50/50 v/v) | 1 mol/L | 10.8 mS/cm | 5.7 V | 982 \$/L$^c$ | [1, 7, 8] |
| **Ionic liquids** | | | | | |
| 1-Ethyl-3-methylimidazolium bis-(trifluoromethylsulfonyl)-imide (EMITFSI) | 3.9 mol/L | 9 mS/cm | 4.5 V | 198 \$/g$^b$ | [9] |
| 1-Butyl-1-methylpyrrolidinium bis(trifluoromethanesulfonyl)imide (PYR$_{14}$TFSI) | 6 mol/L | 2.8 mS/cm | 6.6V | 22 \$/g$^b$ | [5] |
| **Salts used in WiSE** | | | | | |
| Lithium Bis(trifluoromethane)sulfonimide (LiTFSI) | 21 mol/kg | 10 mS/cm | 3 V | 7 \$/g$^b$ | [10] |
| Potassium acetate (KOAC) | 30 mol/kg | 25 mS/cm | 3.2 V | 0.35 \$/g | [11] |
| Lithium acetate (LiOC) + KOAC | 32 mol/kg KOAc 8 mol/kg LiOAc | 5.3 mS/cm | 2.7 V | 0.9 \$/g + 0.35 \$/g | [12] |
| Sodium perchlorate | 17 mol/kg | 64 mS/cm | 2.8V | 0.33 \$/g | [13] |
| Ammonium acetate (AmAc) | 26.4 mol/kg | 49 mS/cm | 2.9V - 3.3V | 1.2 \$/g | This work* |

$^a$ solvent cost, $^b$ salt cost, $^c$ solution costs from Sigma Aldrich; * using different electrodes.

**Table 2.** Acronym of the AmAc solutions investigated with the corresponding values of molality, salt to solvent molar ratio, molarity, density calculated by eq.(1), experimental and from MD simulations, and excess molar volume (EV) obtained as described in the supplementary information).

| Code | Molality (mol/kg) | AmAc:$H_2O$ Molar ratio | Molarity (mol/L) | Density (kg/L) | | | EV (mL/mol) |
|------|------|------|------|------|------|------|------|
| | | | | Calculated | Exp | MD | |
| 1m | 1 | 1.8:100 | 0.95 | 1.010 | 1.02 | 1.02 | -0.25 |
| 5m | 5 | 9:100 | 3.79 | 1.042 | 1.05 | 1.09 | -0.18 |
| 10m | 10 | 1.8:10 | 6,06 | 1.067 | 1.07 | 1.13 | -0.07 |
| 15m | 15 | 2.7:10 | 7.5 | 1.084 | 1.08 | 1.15 | 0.01 |
| 20m | 20 | 3.6:10 | 8.58 | 1.096 | 1.09 | 1.16 | 0.16 |
| 26.4m | 26.4 | 4.8:10 | 9.57 | 1.107 | 1.10 | 1.17 | 0.23 |
| 30m | 30 | 5.4:10 | 10.10 | 1.112 | 1.11 | 1.18 | 0.24 |

**Table 3.** Diffusion constants (D) for $H_2O$, acetate (AcO$^-$) and ammonium (NH$_4^+$) from MD simulations of the AmAc solutions.

| Solution Code | AmAc:$H_2O$ Molar ratio | D ($10^5$, cm$^2$ s$^{-1}$) | | |
|------|------|------|------|------|
| | | $H_2O$ | AcO$^-$ | NH$_4^+$ |
| 1 m | 1.8:100 | 1.98 | 1.08 | 1.07 |
| 5 m | 9:100 | 1.17 | 0.27 | 0.38 |
| 10 m | 1.8:10 | 0.60 | 0.13 | 0.17 |
| 15 m | 2.7:10 | 0.37 | 0.08 | 0.08 |
| 20 m | 3.6:10 | 0.23 | 0.02 | 0.04 |
| 26.4 m | 4.8:10 | 0.17 | 0.02 | 0.02 |
| 30 m | 5.4:10 | 0.11 | 0.01 | 0.02 |

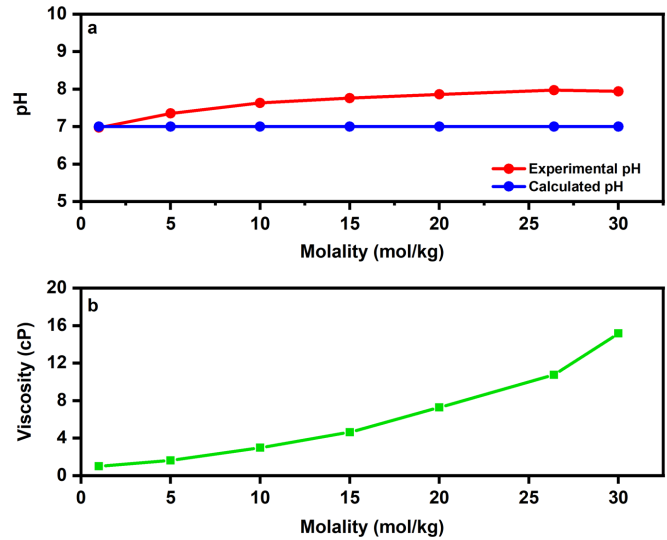

**Figure 1.** a) pH and b) viscosity of AmAC solutions with different molality at room temperature.

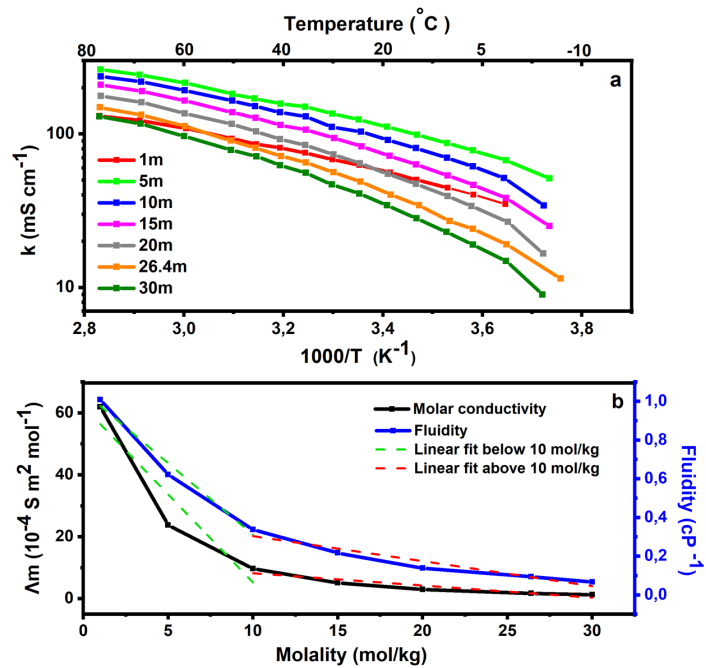

**Figure 2.** Arrhenius plots build on the basis of the conductivity (κ) at different temperatures of the different solutions, b) Trend of the molar conductivity (Λm) and fluidity versus molality at room temperature.

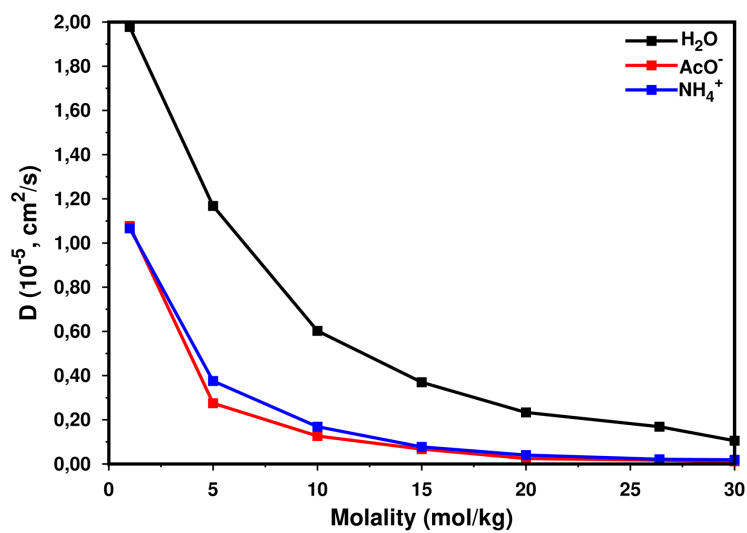

**Figure 3.** Trend of the diffusion constants of $H_2O$, $AcO^-$ and $NH_4^+$ vs. molality.

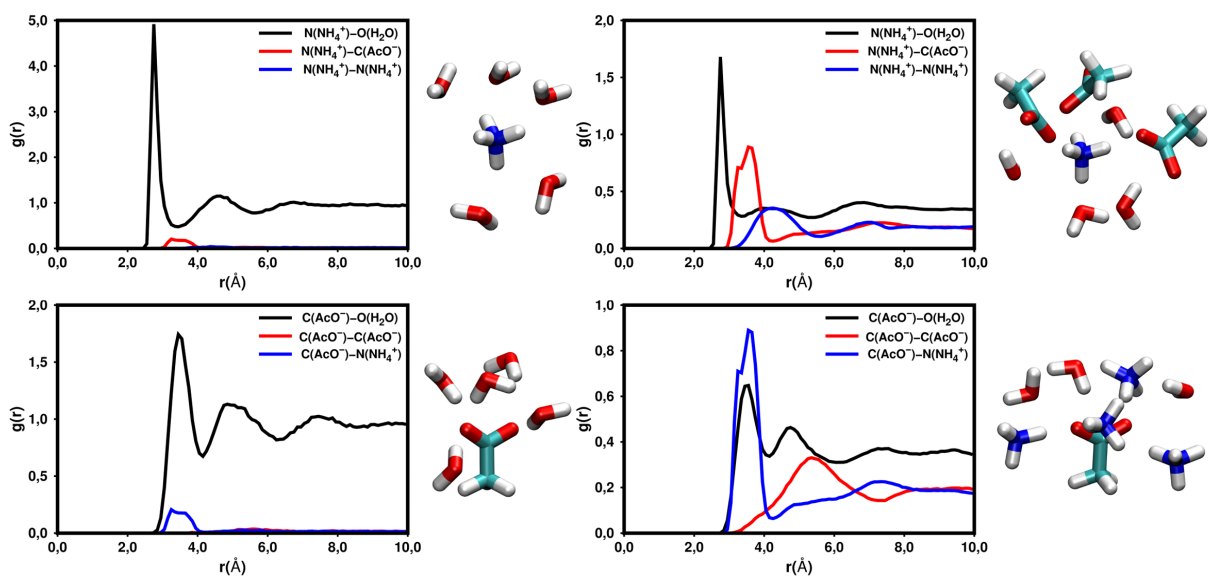

**Figure 4.** Radial distribution function for a 1m mixture (left) and 30 m mixture (right) centred on ammonium cation (top) and acetate anion (bottom). A representative cluster structure is reported.

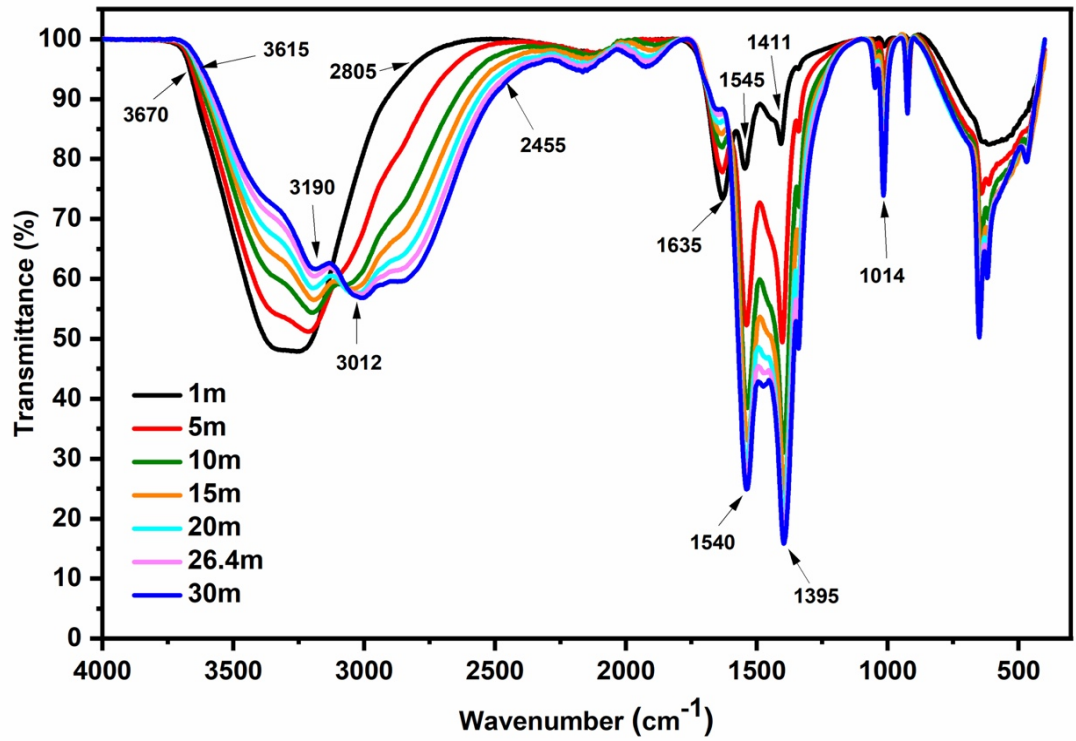

**Figure 5.** IR spectra superimposed of all the studied solutions.

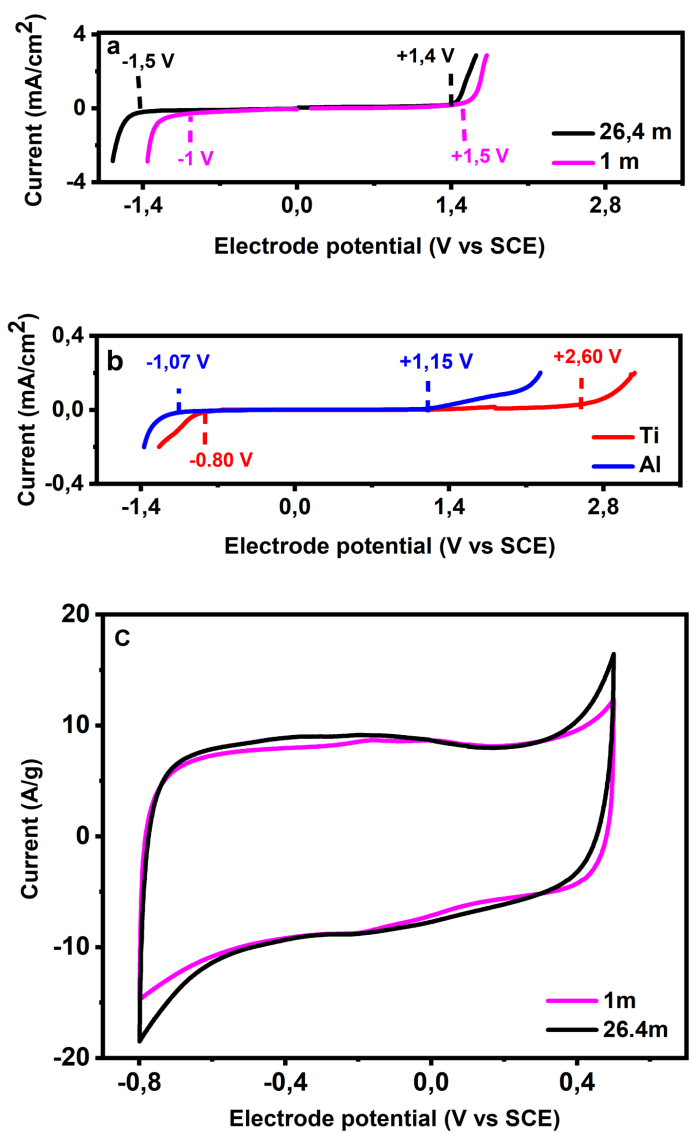

**Figure 6.** Linear sweep voltammetry at 20 mVs$^{-1}$ carried out with (a) glassy carbon electrode (GC) in 1m and 26.4 m solutions, (b) CG, titanium grid and aluminum paper in 26.4 m solution, and (c) CG coated by 0.025 mg Argan shells derived carbon (80% ARG-K-PM, 10% acetylene black, 10% Nafion binder) in 1m and 26.4 m AmAc.

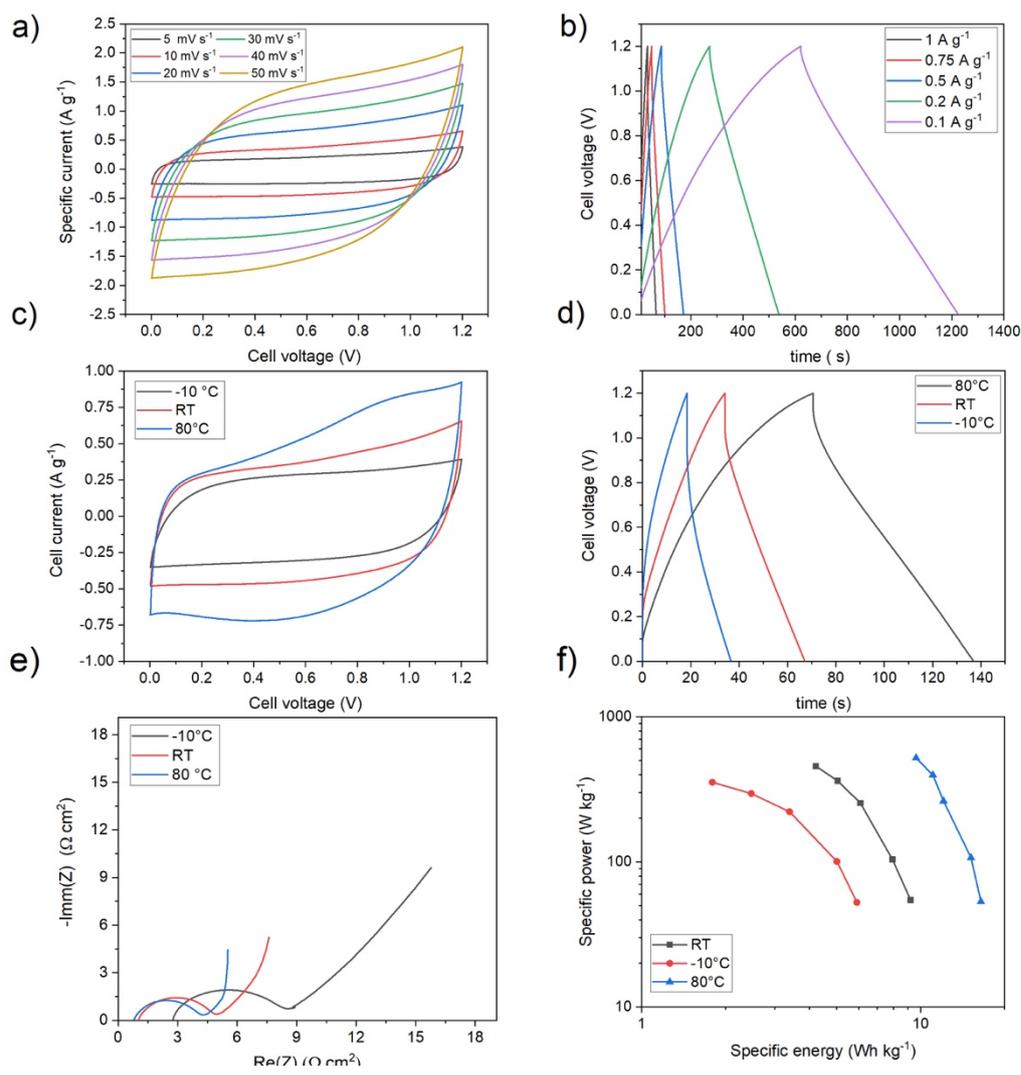

**Figure 7**. Electrochemical test of the asymmetric supercapacitor assembled with ARG-K-PM electrodes and 26.4 m AmAc WiSE: a) CV of the assembled device at scan rate from 5 to 50 mVs and b) GCD at current densities from 0.1 to 1 Ag$^{-1}$ (calculated on the basis of positive and negative electrode mass) at room temperature; c) CV at scan rate 10 mVs$^{-1}$, d) , d) GCD at current densities from 0.1 Ag$^{-1}$ , e) Nyquist plot within a frequency range from 100 KHz to 10 mHz, and f) Ragone plots of the supercapacitor at different temperatures.

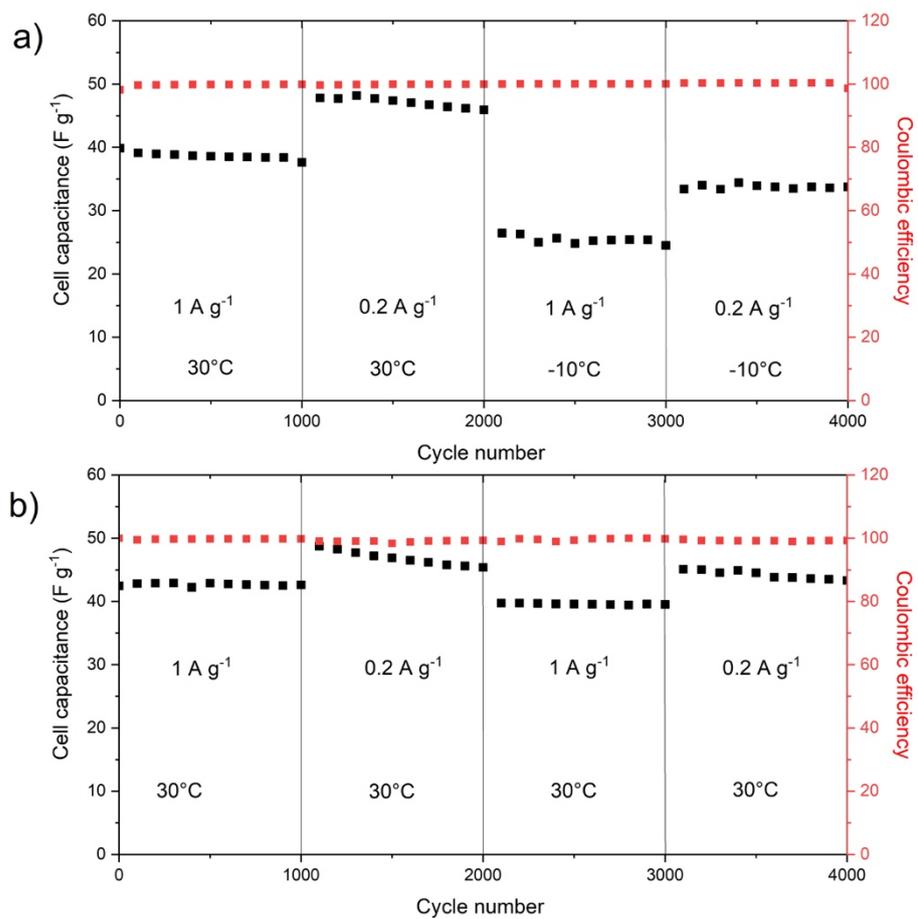

**Figure 8**. Trend of the capacitance under galvanostatic cycling at 0.1A g$^{-1}$ and 1 A g$^{-1}$ of a) the asymmetric supercapacitor with 26.4 m AmAc at RT and -10°C and (b) of the asymmetric capacitor with 1m AmAc at RT.

# Circumneutral concentrated ammonium acetate solution as water-in-salt electrolyte

Mohammad Said El Halimi[a,c], Federico Poli[a], Nicola Mancuso[a], Alessandro Olivieri[a],

Edoardo Jun Mattioli[b], Matteo Calvaresi[b], Tarik Chafik[c], Alberto Zanelli[d],

Francesca Soavi*[a,d]

[a] Laboratory of Electrochemistry of Materials for Energetics, Department of Chemistry "Giacomo Ciamician", Alma Mater Studiorum Università di Bologna, Bologna 40126, Italy

[b] NanoBio Interface Laboratory, Department of Chemistry "Giacomo Ciamician", Alma Mater Studiorum Università di Bologna, Bologna 40126, Italy

[c] Chemical Engineering and Ressources Valorization, Faculty of Sciences and Techniques of Tangier, Abdelmalek Essaadi University, Tangier 90000, Morocco[d] Istituto per la Sintesi Organica e la Fotoreattività (ISOF), Consiglio Nazionale delle Ricerche (CNR), Bologna 40129, Italy

*Corresponding author: francesca.soavi@unibo.it

# Supplementary information

1. **Density calculation**

The theoretical density was calculated by the following equation:

$$d_{calculated} = \frac{m_{AmAc} + m_{H_2O}}{V_{AmAc} + V_{H_2O}} \quad (1)$$

Where $m_{AmAc}, m_{H_2O}$, $V_{AmAc}$ and $V_{H2O}$ are mass and volume of ammonium acetate and water. Using the definition of the salt molality ($C_{AmAc}$) and density ($d_{AmAc}$ = 1.17 kg L$^{-1}$), eq. (1) can be rearranged as in eq (2)

$$d_{calculated} = \frac{(C_{AmAc} * m_{H_2O} * MM_{AmAc}) + m_{H_2O}}{\frac{(C_{AmAc} * m_{H_2O} * MM_{AmAc})}{d_{AmAc}} + V_{H_2O}} = \frac{(C_{AmAc} * MM_{AmAc} * 10^{-3}) + 1}{\frac{(C_{AmAc} * MM_{AmAc} * 10^{-3})}{d_{AmAc}} + d_{H_2O}} \quad (2)$$

where $MM_{AmAc}$ is the ammonium acetate molar mass (77.08 g mol$^{-1}$) and $d_{H2O}$ is the density of water (1 kg L$^{-1}$).

## 2. Excess molar volume EV calculation

EV has been calculated using the following equation:

$$EV = MM_{AmAc} X_{AmAc} \left( \frac{1}{d_{solution}} - \frac{1}{d_{AmAC}} \right) + MM_{H_2O} X_{H_2O} \left( \frac{1}{d_{solution}} - \frac{1}{d_{H_2O}} \right) \qquad (3)$$

where $X_{AmAc}$ and $X_{H2O}$ are the mole fractions of ammonium acetate and water; $d_{solution}$ is the measured solution density (experimental density values from Table 2).

## 3. pH calculation

The salt $CH_3COONH_4$ (AmAc) is composed by weak acidic and base ions that in water give the following equilibrium

$$CH_3COO^- + NH_4^+ + H_2O \leftrightarrow CH_3COOH + NH_4OH \qquad (4)$$

with the equilibrium constant

$$K_{eq} = [CH_3COOH][NH_4OH] / [CH_3COO^-][NH_4^+] \qquad (5)$$

The acid constant of the acetic acid is

$$K_a = [CH_3COO^-][H^+] / [CH_3COOH] = 1.75 \times 10^{-5} \quad \text{(at room temperature)} \qquad (6)$$

The base constant of ammonium hydroxide is

$$K_b = [NH_4^+][OH^-] / [NH_4OH] = 1.74 \times 10^{-5} \quad \text{(at room temperature)} \qquad (7)$$

By combining eqs 5, 6, and 7, $K_{eq}$ can be rewritten as:

$$K_{eq} = \frac{([CH_3COO^-][H^+]/K_a)([NH_4^+][OH^-]/K_b)}{[CH_3COO^-][NH_4^+]} = \frac{[H^+][OH^-]}{K_a K_b} = \frac{K_w}{K_a K_b} \qquad (8)$$

From the stoichiometry of ammonium acetate :

$$[CH_3COO^-] = [NH_4^+] \quad \text{and} \quad [CH_3COOH] = [NH_4OH] \qquad (9)$$

then

$$K_{eq} = [CH_3COOH]^2 / [CH_3COO^-]^2 = K_w / (K_a K_b) \qquad (10)$$

From the acetic acid dissociation equilibrium (6):

$$[CH_3COOH]/[CH_3COO^-] = [H^+]/K_a \qquad (11)$$

Rewriting the expression for $K_{eq}$,

$$K_{eq} = ([H^+]/K_a)^2 = K_w/(K_a K_b) \qquad (12)$$

which yields the formula

$$[H^+] = \sqrt{(K_w K_a / K_b)} \qquad (13)$$

i.e. for AmAC

$$[H^+] = \sqrt{(10^{-14} \times 1.75 \times 10^{-5}/1.75 \times 10^{-4})} = 10^{-7} \text{ mol/L}$$

(14)

hence, pH= -log [H$^+$] = 7

### 4. Ionic conductivity and temperature dependence (Arrhenius behaviors)

Arrhenius behaviors can be described by the equation below, where the logarithm of the specific conductivity (κ) linearly decreases with the reciprocal of the temperature:

$$\log \kappa = \log A - \frac{E_a}{RT} \tag{15}$$

where A is a constant (according to the collision theory, A is the frequency of collisions in the correct orientation), $E_a$ is the activation energy, R is the universal gas constant and T is the temperature.

### 5. Symmetric supercapacitor with c two identical ARG-K-PM electrodes and 26.4 m AmAc WiSE

The highest charge cut-off voltage of the symmetric supercapacitor that enabled high coulombic efficiency (> 99 %) was 0.8 V.

Taking into account the good conductivity response at low and high temperatures of 26.4 m AmAc, we carried out a preliminary study of the supercapacitor performance at room temperature (RT), -10°C and 80°C. Fig. S1 reports the corresponding CV, GCD, EIS and Ragone plots. The highest charge cut-off voltage that enabled high coulombic efficiency (> 99 %) was 0.8 V.

Fig. S1a shows the CVs at RT carried out with increasing the scan rate from 5 to 50 mV s$^{-1}$ the curves exhibit a quasi-rectangular shape profile demonstrating a good capacitive behaviors of the electrodes. However, at high scan rate the rectangular characteristic of the curves becomes less remarkable indicating that the capacitive behaviors of the electrodes decrease due to the incomplete diffusion of electrolyte ions into the porous structure of material electrodes.

The GCD was performed at current density ranging from 0.1 Ag$^{-1}$ to 1 Ag$^{-1}$. The GCD profiles at room temperature are reported Fig. S1b. They exhibit triangular shape indicating a good reversibility and capacitive behavior of the device. Also all GCD curves show a small ohmic drop, therefore suggesting a low ESR. Figures S1c, S1d, S1e compare the CVs (at 10 mV s$^{-1}$), the GCD profiles (at 0.1 Ag$^{-1}$) and the Nyquist plots (100 kHz - 10 mHz frequency range) collected at -10°C, RT and 80 °C. In Fig. S1c all the CV curves exhibit rectangular shape, as well as in Fig. S1d all GCD keep a triangular shape at all the temperatures, therefore indicating a good electrochemical capacitive behavior of the device both at extreme temperatures. The calculated gravimetric supercapacitor capacitances from GCD curve at 0.1 Ag$^{-1}$ were 34 Fg$^{-1}$, 47 Fg$^{-1}$ and 72 Fg$^{-1}$ at -10° C, RT and 80°C. These values correspond to electrode specific capacitance values of 136 F g$^{-1}$, 188 F g$^{-1}$, and 288 F g$^{-1}$. The maximum energy densities $E_{max}$ are 3 Wh kg$^{-1}$ (-10°C), 4 Wh kg$^{-1}$ (RT) and 6 Wh kg$^{-1}$ (80°C). The ESR values evaluated by the ohmic drop at the beginning of discharge resulted in ca. 7.7 cm$^2$ at -10°C, 2.9 cm$^2$ at RT, and 2.3 cm$^2$ at 80°C. These values well compare with the low frequency resistances of the cells shown by the Nyquist plots reported in Fig.S1e. It is worth noting the low ESR exhibited by the cells even at the lowest temperature. The plots indicate that the decrease of temperature mainly impacts on ion diffusion in the porous electrode architecture (low frequency tail of the Nyquist plots). On the other hand, MD simulation and experimental data reported in the previous sections already indicated that cooperative mechanisms are responsible for AmAc WiSE ion conductivity. In turn, this impacts on the kinetics of the electrical double layer formation at the electrode/electrolyte interface, especially at the lowest temperatures. From ESR, maximum power densities $P_{max}$ of 0.7 kW kg$^{-1}$ (-10°C), 2 kW kg$^{-1}$ (RT), and 2.6 kW kg$^{-1}$ (80°C) were measured.

The practical specific energy and power delivered by the supercapacitor at different currents and temperatures are compared in the Ragone plot reported in Figure S1f. The maximum specific energy at the lowest current, and the maximum power is delivered at the highest current. At 0.1 A g$^{-1}$, the specific energy is 2.3 Wh kg$^{-1}$ (-10°C), 3.3 Wh kg$^{-1}$ (RT), and 5.0 Wh kg$^{-1}$ (80°C). At 4 A g$^{-1}$, the specific power is 215 W kg$^{-1}$ (-10°C), 360 W kg$^{-1}$(RT), and 415 W kg$^{-1}$ (80°C).

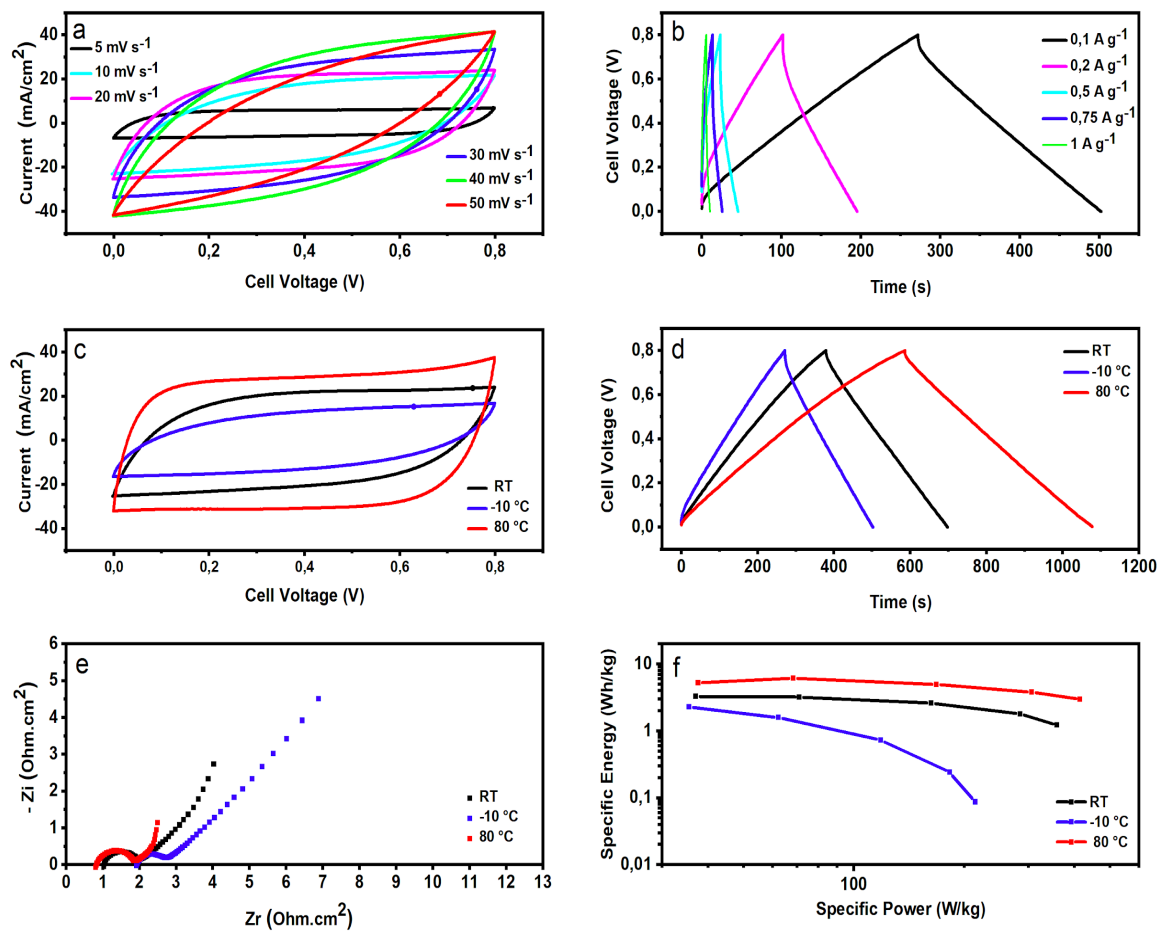

Figure S1. Electrochemical test of the symmetric supercapacitor assembled with ARG-K-PM electrodes and 26.4 m AmAc WiSE: a) CV of the assembled device at scan rate from 5 to 50 mVs$^{-1}$ and b) GCD at current densities from 0.1 to 1 Ag$^{-1}$ (calculated on the basis of positive and negative electrode mass) at room temperature; c) CV at scan rate 10 mVs$^{-1}$, d) GCD at current densities from 0.1 Ag$^{-1}$, e) Nyquist plot within a frequency range from 100 KHz to 10 mHz, and f) Ragone plots of the supercapacitor at different temperatures.

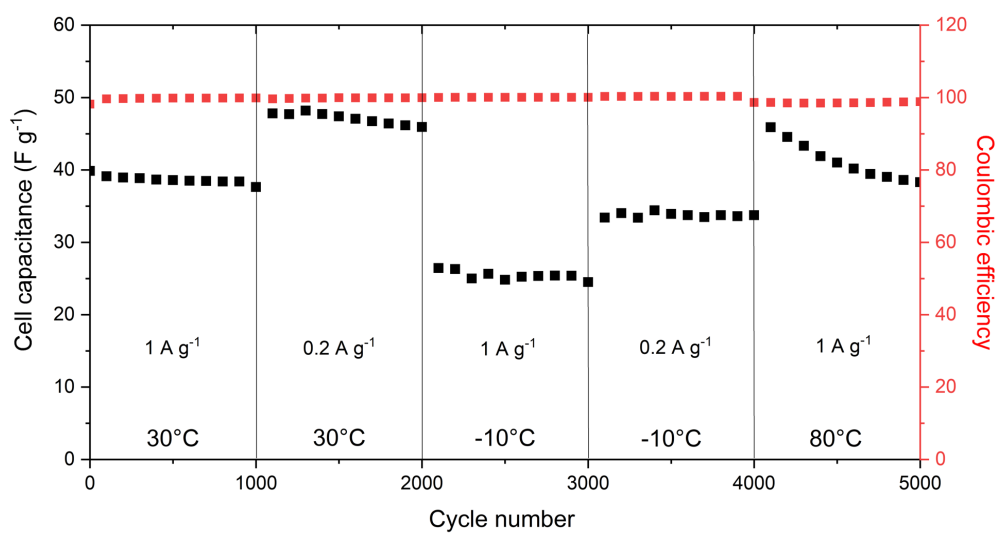

Figure S2. Trend of the capacitance under galvanostatic cycling at 0.1A g$^{-1}$ and 1 A g$^{-1}$ of the asymmetric supercapacitor with 26.4 m AmAc at RT and -10°C (with V$_{max}$ of 1.2 V), and 80°C (with V$_{max}$ of 1V).